# Structural relaxation of E'$_g$ centers in amorphous silica

S. Agnello, R. Boscaino, G. Buscarino, M. Cannas and F.M. Gelardi

Istituto Nazionale per la Fisica della Materia and Department of Physical and Astronomical Sciences, University of Palermo, Via Archirafi 36, I-90123 Palermo, Italy

(Receipt date)

We report experimental evidence of the existence of two variants of the E'$_\gamma$ centers induced in silica by $\gamma$ rays at room temperature. The two variants are distinguishable by the fine features of their line shapes in paramagnetic resonance spectra. These features suggest that the two E'$_\gamma$ differ for their topology. We find a thermally induced interconversion between the centers with an activation energy of about 34 meV. Hints are also found for the existence of a structural configuration of minimum energy and of a metastable state.

The importance of point defects in silicon dioxide (SiO$_2$) both in crystalline and amorphous (a-SiO$_2$) polymorphs has been shown in connection with the use of this material in various optical and electronic devices [1-2]. Among these point defects one of the most diffusely characterized and discussed is the E'$_\gamma$ center in a-SiO$_2$ that is an intrinsic paramagnetic point defect, i.e. involves only Si and O atoms [3-5]. Despite the huge number of theoretical and experimental works, the structure and the spectroscopic properties of this defect are still widely debated [1-2, 6]. Recently, a new model for the E'$_\gamma$ center has been suggested by Uchino et al. [7-8] in which the defect originates from an "edge sharing" oxygen vacancy (triangular oxygen-deficiency center, TODC) by trapping a hole (bridged hole-trapping oxygen-deficiency center, BHODC): =Si$^\bullet$-O-$^+$Si= (where = represents bonds to two distinct O atoms, $\bullet$ is an unpaired electron and + is the trapped hole). This model succeeds to account for the experimental value, 42 mT, of the strong $^{29}$Si hyperfine splitting [5, 9]. Moreover, due to the absence of any structural constraint, it seems suitable to describe the structure of the defect in a-SiO$_2$. The model by Uchino et al. is rather different from the "historical" one originally proposed by Feigl-Fowler-Yip (FFY) [10-11] for the E'$_1$ center [3] in $\alpha$-quartz, consisting in a hole trapped by a neutral oxygen monovacancy: ≡Si-Si≡ +h$^+$ → ≡Si$^\bullet$ + $^+$Si≡ (where ≡ represents the bonds with three distinct O atoms). Also this model, refined up to the puckered configuration, gave fair agreement both with the strong and the weak hyperfine splittings measured in $\alpha$-quartz [12]. In the latter configuration, one assumes that the $^+$Si≡ group relaxes backward away from the vacancy and the Si$^+$ is also bonded to a normal bridging O becoming again fourfold coordinated [12, see Fig.1 in Ref.8]. Even though the presence of suitably positioned bridging O in the disordered matrix is unknown, the FFY model has been extended also to the E'$_\gamma$ center in a-SiO$_2$, on the basis of the close analogies of its electron paramagnetic resonance (EPR) features to those of the E'$_1$ [13]. This FFY model has been supported by various theoretical calculations [14-17], however, Uchino et al., in several unconstrained large clusters representative of a-SiO$_2$, found no backward puckering and that the unpaired electron becomes equally shared between the two adjacent Si [8].

A closely related aspect of this debate regards the E'$_\beta$ center of a-SiO$_2$ [1, 18] or its $\alpha$-quartz equivalent E'$_2$ [19]. Griscom and Rudra et al. proposed a monovacancy model in which one Si bonds a H atom and the other has the unpaired electron that points its orbital away from the vacancy direction in a kind of void after a large relaxation: ≡Si-H + ≡Si$^\bullet$. At variance, Uchino et al. proposed that E'$_\beta$ arises from the TODC when one Si is bonded to a H atom and the other one holds the unpaired electron: =Si$^\bullet$-O-$^{\text{H}^-}$Si= [8].

It is worth to note that the main EPR features (g-values and hyperfine constants) of the E' center are theoretically explained merely in terms of the ≡Si$^\bullet$ structure. However, Griscom evidenced various typologies of E' centers in a-SiO$_2$, distinguished by small variations of the EPR line shape that can be ascribed to different atomic compositions of the neighborhood of the unpaired electron [1, 18]. Moreover, just for the E'$_\gamma$ center, an evolution of the line shape following thermal treatments at the temperature of T~500 K was found [18, 20], so revealing the existence of unexplained degrees of structural freedom of the defect.

The above considerations indicate that the E'$_\gamma$ center deserves further experimental investigation and, to better evidence its degrees of freedom, a fine study of the EPR line shape. We report here a detailed EPR study of the E'$_\gamma$ centers induced by γ irradiation in a variety of commercial a-SiO$_2$ that can be grouped as follows [21]: *Natural dry* - Infrasil 301 (I301), EQ 906 (EQ906), EQ 912 (EQ912), Puropsil A (QPA); *Natural wet* - Herasil 1 (H1), Herasil 3 (H3), Homosil (HM); *Synthetic dry* - Suprasil 300 (S300); (EQ906, EQ912 and QPA supplied by Quartz&Silice all the others by Heraeus). Each sample is slab shaped with size 5 x 5x 1 mm$^3$. Different pieces of each material were exposed to γ rays at room temperature in a $^{60}$Co source, accumulating doses, D, in the range from $10^{-1}$ kGy to $10^4$ kGy at the rate ~7 kGy/hr. EPR measurements were carried out at room temperature with a Bruker EMX spectrometer working at frequency ν ≈ 9.8 GHz in the first derivative mode. E'$_\gamma$ centers spectra were taken at modulation magnetic field frequency of 100 kHz, modulation amplitude of 0.01 mT and at microwave power of 800 nW; the latter two conditions avoid line shape distortions. The main spectroscopic g-values were determined by accurate frequency measurements allowing us to find the differences between the g's with a maximum error of ± 0.00001. The spin concentration, $C_s$, of one sample of each material was determined, with absolute accuracy of 20 %, using the instantaneous diffusion method in spin echo decay measurements carried out in a pulsed spectrometer [22]. For the other samples, $C_s$ was evaluated, with an accuracy of 10%, by comparing the double integral of the EPR spectrum with that of the reference sample. Our $C_s$ detection limit is estimated to be ~$10^{15}$ spins/cm$^3$.

γ-irradiation induces the E'$_\gamma$ centers in all the investigated materials. They begin to be detectable at doses that strongly depend on the material. A typical dose-dependence is reported in the inset of Fig.1a for I301 [23-24]: $C_s$ initially grows linearly and then reaches a constant value maintained up to the highest doses. This feature is evidence of a generation process from precursors [25]. A direct activation of the matrix has been observed at doses higher then those considered here [24, 25]. Together with the variation of $C_s$, we observed a modification of the EPR line shape of the E'$_\gamma$ centers on increasing the dose D. In Fig.1a, we report the line shapes as detected in the I301 material after γ doses of 0.5, 50 and 5000 kGy. We note a gradual shift of zero-cross point towards smaller resonance fields and an overall broadening of the line shape, on increasing the dose from 0.5 kGy upwards. A quantitative analysis can be carried out looking at the principal g-values, $g_1$, $g_2$ and $g_3$, experimentally determined from the field values at which the first maximum ($g_1$), the zero-crossing point ($g_2$), and the minimum of the EPR spectra ($g_3$) occur [26]. The difference $\Delta g_{1,2} = g_1 - g_2$ varies from 0.00124 at D = 0.5 kGy (low-dose limit) up to 0.00115 at D =5000 kGy (high-dose limit). The variation of $\Delta g_{1,3} = g_1 - g_3$ is less pronounced on increasing the dose, $\Delta g_{1,3} = 0.00147$ at D= 0.5 kGy and $\Delta g_{1,3} = 0.00142$ at D =5000 kGy. These gradual line shape variations occur between 10 kGy and 1000 kGy.

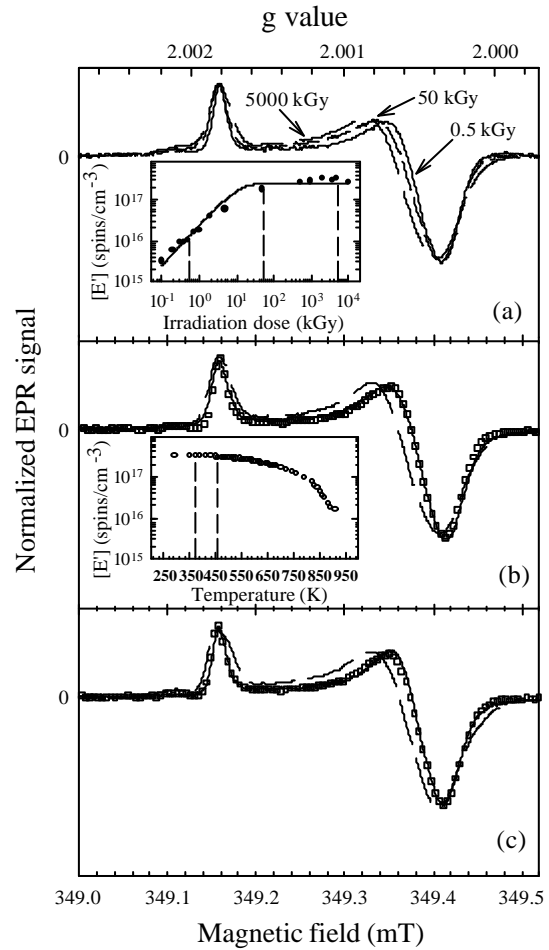

FIGURE. 1. EPR spectra of the E'$_\gamma$ centers normalized to the peak to peak amplitude and horizontally shifted to overlap at the first maximum. (a) I301 samples irradiated at doses 0.5 kGy (solid line), 50 kGy (short-dashed) and 5000 kGy (long-dashed); in the inset the E'$_\gamma$ concentration is reported as a function of the dose (the solid line is a guide to the eye). (b) I301 sample after irradiation at 4000 kGy (dashed) and after isochronal thermal treatments up to T = 460 K (solid), the squares refer to the reference sample I301 irradiated at 0.5 kGy; in the inset the E'$_\gamma$ concentration is reported as a function of the temperature in the isochronal thermal treatments. (c) S300 sample irradiated at $10^4$ kGy (dashed) and after 9 hours of thermal treatment at T = 500 K (solid), the squares refer to the reference sample I301 irradiated at 0.5 kGy.

The line shape reported for the 0.5 kGy irradiated sample is characteristic of the low dose region whereas that observed at 5000 kGy is peculiar of the high dose. For convenience, hereafter we adopt the symbols L1 and L2 for the low and high dose line shapes, respectively. We note that $\Delta g_{1,2}$ and $\Delta g_{1,3}$ found for L2 are in strict agreement with those reported by Griscom for the E'$_\gamma$; as suggested by the simulated spectra [18] L2 could be related to an orthorhombic symmetry whereas L1 to an axial one. Moreover, we guess that the broadening of g-values distribution passing from L1 to L2 is due to irradiation increased vitreous disorder. The phenomenology just reported for the I301 is a feature common to all the other materials. Indeed, we observe the line shape L1 after low γ-doses, with the uncertainty of 0.001 mT, and the same line shape L2 after high γ-doses in all the silica types considered here. It is worth to note that we have verified that the variation from L1 to L2 is not related to the concentration of centers, i.e. to dipole-dipole interaction [4]. As an example, the line shape L2 is observed also in synthetic wet material (not reported here) at a concentration of $4 \times 10^{15}$ spins/cm$^3$ whereas at the same concentration of E'$_\gamma$ centers the line shape L1 is observed in both natural and synthetic dry materials. We can infer that these variations of the line shape are intrinsic to the process of defect generation as they occur in all the materials.

To further investigate the line shape variation we carried out a series of thermal treatments in the sample I301 previously irradiated at 4000 kGy in which we recorded the line shape L2 and $C_s \approx 3.3 \times 10^{17}$ spins/cm$^3$. Various isochronal treatments with time fixed to 25 min were carried out at normal atmosphere in the temperature range 350 K≤T≤910 K in an electric furnace with T stabilized within ±3 K. After each treatment the sample returned to room temperature before EPR measurements. For T< 370 K no significant variation occurs in the line shape nor in $C_s$, as shown in the inset of Fig.1b. A gradual change from L2 toward L1 occurs for 370 K≤T≤ 460 K. Actually, as shown in Fig. 1b, the line shape after the treatment at 460 K coincides with that in the same material exposed to a dose of 0.5 kGy. A noteworthy aspect is that after these treatments $C_s$ is $2.9 \times 10^{17}$ spins/cm$^3$ indicating that only a very low quantity of centers has been destroyed and ruling out definitively that the line shape changes are due to dipole-dipole interaction. At higher temperature we observe an additional reduction of $C_s$ down to $2.7 \times 10^{17}$ spins/cm$^3$ and a very small variation of the line shape; at T = 520 K we found $\Delta g_{1,2} = 0.00125$ and $\Delta g_{1,3} = 0.00147$, very close to L1. Finally, on increasing further the temperature only a reduction of $C_s$ occurs and temperatures as high as 800 K are required to obtain its large decrease. Similar results were obtained as well in all the investigated materials.

Here we limit ourselves to report on a somewhat different thermal treatment carried out in the synthetic dry material S300. After γ-irradiation at the dose of $10^4$ kGy, we measured $C_s \approx 1.1 \times 10^{17}$ spins/cm$^3$ and a line shape L2, as shown in Fig.1c. The sample was heated for nearly 9 hr at 500 K in a He-filled dewar. So long a heat treatment caused only a slight decrease of $C_s$ down to $7 \times 10^{16}$ spins/cm$^3$ but an evident variation of the line shape from L2 to L1. This result is relevant as the very high purity of synthetic material with respect to the natural one rules out the possibility that the line shape modification can be ascribed to impurity-related effects.

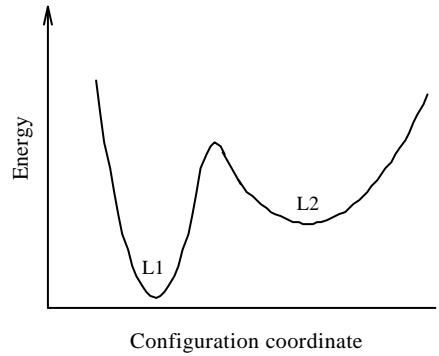

FIGURE 2. Total energy for the structural configurations of the E'$_\gamma$ centers related to the experimental line shapes L1 and L2.

Results reported above suggest that the E'$_\gamma$ center possesses a structural configuration of minimum energy and a metastable state. An axial line shape L1 is observed after low-dose irradiation and can be associated to those centers whose formation is energetically favored. On increasing the dose other centers are formed in a metastable configuration state associated to an orthorhombic line shape L2. A conversion from L2 to L1 is induced by warming and can be explained considering that some thermal vibration energy is employed in a structural conversion where the metastable centers switch to the more stable ones. A qualitative representation of these features is outlined in Fig.2 where the energy related to a given configuration is reported as a function of a generic configuration coordinate. The energy well associated to the line shape L1 is expected to be lower and narrower with respect to that of L2, explaining both the major energy stability and the minor broadening of g-values distribution experimentally detected. An energy barrier between the two wells, corresponding to a thermal energy of ~34 meV (T~400 K), separates the two configurations thus explaining the thermally activated conversion.

Now we try to interpret our results in terms of the existing models of the E'$_\gamma$ centers. We consider the asymmetrically relaxed oxygen monovacancy [12]. In this model the unpaired electron points toward the vacancy so that the observed variation of the line shape should be attributed to the perturbative role of $Si^+$ [18, 27]. In particular, the L1 center should result from backward puckering of $Si^+$ which bonds to a normal bridging O, being stabilized and energetically favored. L2 would be the unpuckered $E'_\gamma$ center, $Si^+$ being the origin of the orthorhombic character. The above scheme seems suitable to explain our results, however it presents some faults. Indeed, theoretical calculations predict only one energy minimum for the puckered state and a non puckered state with the unpaired electron shared between the two Si [14]. This structure requires a consistent line shape variation and differences in the hyperfine structures not observed. Moreover, the conversion from monovacancy to $E'_\gamma$ center is questioned [6] and has not been observed in our samples [28].

In view of this partial failure, we wish to put forward an alternative explanation for our results. We tentatively assign the configuration L1 of the $\equiv Si^\bullet$ moiety to an $E'_\beta$-like center [18], in which the bonds of the three basal O with the near neighbors Si are in the same side of the unpaired electron orbital (backward projection). A similar atomic arrangement was proposed for the $E'_2$ in quartz and the $E'_\beta$ (generated at low temperature [18]), to justify a more symmetric line shape. Consequently, we assume that in the configuration L2 these bonds are in the opposite side of the unpaired electron (forward projection). The inter-conversion from L2 to L1 corresponds to $Si^\bullet$ crossing through the plane of the basal O atoms. We note that this movement does not affect the structure of the moiety and is expected to manifest as small variations in the line shape without changing the strong hyperfine structure, as experimentally observed. In this scheme the $E'_\beta$-like structure is energetically favored with respect to the other; in this sense our interpretation deviates from the prediction of the vacancy model where the energy minimum is expected when $Si^\bullet$ is forward projected attracted by $Si^+$. On the other hand, our interpretation seems to be consistent with the Uchino et al. model where the $\equiv Si^\bullet$ moiety should have two energy configurations, corresponding to two different distances from $Si^+$ in BHODC, with a minimum in the backward projection. We note, however, that in this model the g-value perturbation induced by $Si^+$ was not estimated. More theoretical works is required for a quantitative comparison. Further insight could also arise from the study of the optical absorption band at 5.8 eV related to the E' center [29].

We thank D.L. Griscom and M. Leone for very useful discussions, E. Calderaro for taking care of the γ irradiation. This work was financially supported by Ministero Italiano della Ricerca Scientifica.